\documentstyle[epsf]{mn}

\begin{document}
\title{Non-gaussian CMB temperature fluctuations from peculiar 
velocities of clusters}
\author[N. Yoshida, R. K. Sheth, \& A. Diaferio]
{Naoki Yoshida$^1$, Ravi K. Sheth$^2$, \& Antonaldo Diaferio$^3$\\
$^1$ Max-Planck-Institut f\"{u}r Astrophysik, Karl-Schwarzschild-str.1
Garching bei M\"unchen, D85748 Germany\\
$^2$ NASA/Fermilab Astrophysics Group, Batavia, IL 60510-0500\\
$^3$ Dipartimento di Fisica Generale ``Amedeo Avogadro'', 
Universit\`a di Torino, Italy\\
\smallskip
Email: naoki@mpa-garching.mpg.de, sheth@fnal.gov, diaferio@ph.unito.it
}
\date{Revised, submitted to MNRAS 2001 August 8}

\maketitle

\begin{abstract}
We use numerical simulations of a (480 Mpc$/h$)$^3$ volume to 
show that the distribution of peak heights in maps of the temperature 
fluctuations from the kinematic and thermal Sunyaev-Zeldovich effects
will be highly non-Gaussian, and very different from the peak height 
distribution of a Gaussian random field.  We then show that it is a 
good approximation to assume that each peak in either SZ effect is 
associated with one and only one dark matter halo.  This allows us to 
use our knowledge of the properties of haloes to estimate the peak
height distributions.  At fixed optical depth, the distribution of 
peak heights due to the kinematic effect is Gaussian, with a width 
which is approximately proportional to optical depth; the 
non-Gaussianity comes from summing over a range of optical depths.  
The optical depth is an increasing function of halo mass, and the 
distribution of halo speeds is Gaussian, with a dispersion which is 
approximately independent of halo mass.  
This means that observations of the kinematic effect can be used to 
put constraints on how the abundance of massive clusters evolves, and 
on the evolution of cluster velocities.  
The non-Gaussianity of the thermal effect, on the other hand, comes 
primarily from the fact that, on average, the effect is larger in more 
massive haloes, and the distribution of halo masses is highly 
non-Gaussian.  We also show that because haloes of the same mass may 
have a range of density and velocity dispersion profiles, the relation 
between halo mass and the amplitude of the thermal effect is not 
deterministic, but has some scatter.  
\end{abstract}


\section{Introduction}
The inverse Compton scattering of free electrons in the intracluster 
plasma with the photons of the cosmic background radiation produce 
secondary fluctuations in background radiation temperature maps. The 
electron motion is due both to thermal motions within the plasma, 
and to coherent bulk motions of the plasma (e.g. Sunyaev \& Zeldovich
1980).  
Simulations suggest that the distribution of these temperature 
fluctuations will be quite non-Gaussian (Seljak, Burwell \& Pen 2001), and that the best hope of 
detecting this effect is in the tails of the distribution 
(e.g. da Silva et al. 2000).  
In this paper, rather than studying the distribution of temperature 
fluctuations in random pixels, we study the distribution of fluctuations 
in pixels which are peaks.  We show that the distribution of these peak 
heights should be highly non-Gaussian, and that the exact shape can be 
computed if the spectrum of initial density fluctuations is known.  
Comparison with estimates of the peak height distribution from large 
$n$-body simulations shows that our model is quite accurate.  
This means that observations of the peaks alone should allow one to 
place constraints on what the initial conditions must have been.  
In this paper, we restrict attention to the simplest case in which 
signals come from a single redshift, chosen to be close to the present 
epoch, when nonlinear structures have already grown. Work in progress 
studies contributions from higher redshifts; even when a range of 
redshifts contribute, our conclusion that the final distribution of 
fluctuations should be non-Gaussian, and related to the initial density 
fluctuation distribution, should remain unchanged.  

This paper is organized as follows.  
Section~\ref{simuls} describes our simulations briefly.
Section~\ref{kszeffect} studies the kinematic effect, and 
Section~\ref{tszeffect} studies the thermal effect.  
The thermal effect is closely related to the optical depth, and so 
this section also includes a study of the optical depth distribution.  
A final section summarizes our findings.  

\section{The simulations}\label{simuls}
The simulation we will use was recently carried out by 
the Virgo Consortium.  It uses $512^3$ particles in a cosmological 
box of $480h^{-1}$Mpc on a side.  The cosmological model is flat with 
matter density $\Omega_{0}=0.3$, cosmological constant 
$\Omega_{\Lambda}=0.7$ and expansion rate at the present time 
$H_{0}=70$km$^{-1}$Mpc$^{-1}$.  It has a CDM initial power spectrum 
computed by CMBFAST (Seljak \& Zaldarriaga 1996), normalized to the 
present abundance of galaxy clusters so that $\sigma_{8}$=0.9.
We emphasize that, if the kinematic effect arises from the peculiar 
velocities of clusters, then a large simulation box such as ours is 
essential for studying this effect:  smaller boxes miss a significant 
fraction of the power which generates velocities, so they are liable 
to underestimate the magnitude of the effect (see Sheth \& Diaferio
2001 for more discussion on the finite box size effect). 
In addition, our large simulation contains a large number of massive 
haloes which smaller simulation boxes are likely to misestimate.  
The proper population of massive halos is essential for studying the 
distribution of the thermal effect (e.g. Refregier \& Teyssier 2000).

From the simulation outputs we create maps of the thermal SZ effect, 
the kinematic SZ effect, and the Thomson optical depth.
We compute the local gas density and velocity from those of dark matter,
and the gas temperature is computed from the local dark
matter velocity dispersion, following the 
procedure outlined in Diaferio, Sunyaev, \& Nusser (2000). 
We project
the simulation 
box on a fine 2400$^{2}$ grid in a random direction; thus, each 
pixel in the grid is $L=200h^{-1}$kpc on a side.  This pixel scale $L$ 
is about an order of magnitude larger than the 30$h^{-1}$kpc spatial 
resolution of the simulation.

\begin{figure}
\centering
\epsfxsize=\hsize\epsffile{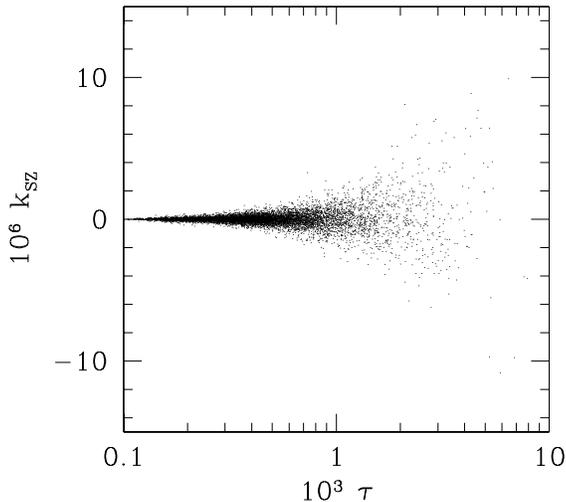}
\caption{Scatter plot showing the height of a $k_{\rm SZ}$ peak 
as a function of optical depth $\tau$.}
\label{scatter}
\end{figure}

\section{The kinematic effect}\label{kszeffect}
The optical depth is defined by 
\begin{equation}
\tau = \sigma_{\rm T} \int n_{\rm electron}(l)\ {\rm d}l 
\approx {\sigma_{\rm T}\over m_{\rm proton}} 
{\Omega_{\rm b}\over\Omega_0}\rho_{\rm crit}
\int {\rm d}l\ {\rho_{\rm dm}(l)\over\rho_{\rm crit}},
\label{taul}
\end{equation}
where $\sigma_{\rm T}$ is the Thomson scattering cross-section, 
$n_{\rm electron}(l)$ is the density of electrons along the line of
sight at $l$, 
$\rho_{\rm dm}(l)$ is the mass density along the line of sight at $l$, 
$\rho_{\rm crit}\Omega_0$ is the average mass density of the background, 
and $\Omega_{\rm b}$ denotes the abundance of baryons.  In what follows 
we have set $\Omega_{\rm b} = 0.0125/h^2$.  The second equality shows 
the approximations used by Diaferio, Sunyaev \& Nusser (2000) to relate 
the properties of the dark matter particles in their simulations to
those of the gas:  the gas traces the dark matter.  

\begin{figure}
\centering
\epsfxsize=\hsize\epsffile{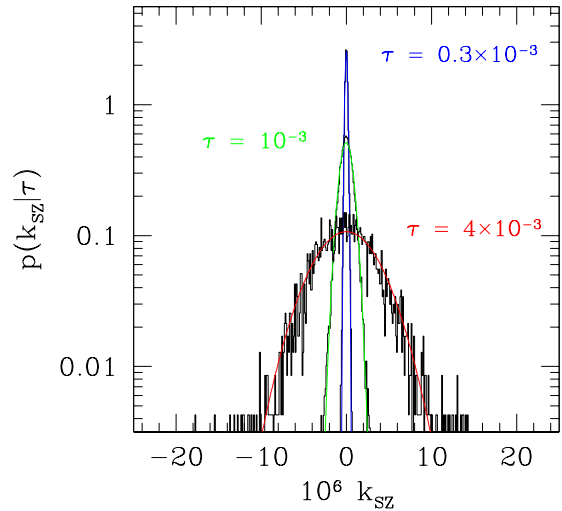}
\caption{Distribution of $k_{\rm SZ}$ peak heights at fixed 
optical depth.  Curves show Gaussians with the same rms.  }
\label{fixedt}
\end{figure}

The kinematic effect temperature fluctuation $k_{\rm SZ}$ is defined by 
\begin{equation}
k_{\rm SZ} = {\Delta T\over T}  
    = {\Omega_{\rm b}\over\Omega_0}{\sigma_{\rm T}\over m_{\rm proton}}
  \rho_{\rm crit}\int {\rm d}l\ {\rho_{\rm dm}(l)\over\rho_{\rm crit}}\,
  {v(l)\over c_{\rm light}},
\end{equation}
where $v(l)/c_{\rm light}$ is the bulk velocity along the line of sight 
at $l$, divided by the speed of light.  Notice that if one thinks of the 
optical depth as a measure of the density, then the kinematic effect is 
a measure of the momentum.  
Below, we first describe what is seen in simulations of the kinematic 
effect, and then we present a model which provides a reasonably good fit 
to the simulations.  

Diaferio et al. (2000) presented a study of peaks in their simulated 
maps of the thermal and kinematic SZ effects.  The panels on the left 
of their Figure~3 show that the mean peak height for the thermal SZ 
effect scales with the square of $\tau$, the optical depth:  
$t_{\rm SZ} \equiv \Delta T(K)/T(K) = -2\tau^2$.  We will consider 
this in the next section.  This section is devoted to a study of the 
relation presented in the panels on the right of their Figure~3.  

\begin{figure}
\centering
\epsfxsize=\hsize\epsffile{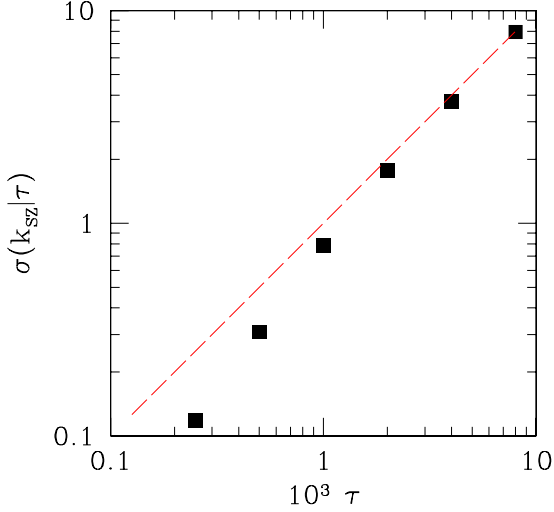}
\caption{The rms value of $k_{\rm SZ}$ at fixed $\tau$ as a function 
of $\tau$.  Dashed line shows the scaling expected if 
$\sigma(k_{\rm SZ}|\tau)$ increases linearly with optical depth. }  
\label{sigmatau}
\end{figure}

\begin{figure}
\centering
\epsfxsize=\hsize\epsffile{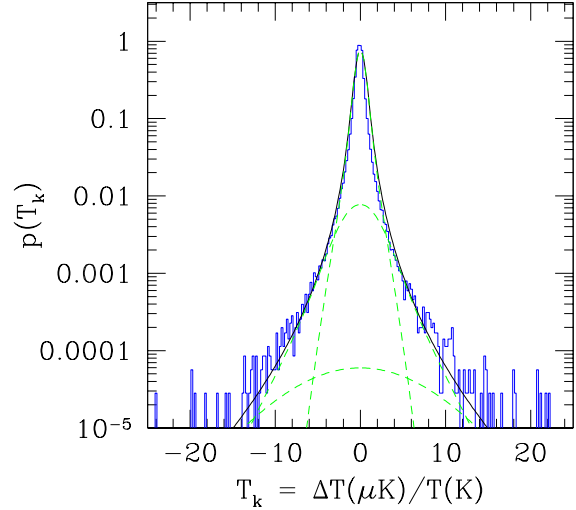}
\caption{Distribution of $k_{\rm SZ}$ fluctuations from sources at 
$z=0$.  Histogram shows the measurements in the $n$-body simulation, 
solid curve shows what our model predicts (eq.~\ref{pkSZ} below).  
Dashed curves show the predicted contribution to the solid curve from
haloes 
with mass in the range $10^{12} - 10^{13}$ (narrowest distribution), 
$10^{13} - 10^{14}$, and $10^{14} - 10^{15}M_\odot/h$ (broadest 
distribution with lowest peak).}
\label{kSZpdf}
\end{figure}

Fig.~\ref{scatter} shows our version of their Figure~3: it shows the 
heights of the peaks in the kinematic effect, $k_{\rm SZ}$, as a 
function of optical depth.  It was constructed following the procedures 
Diaferio et al. outlined; the only difference is that here we use a
$\sim$40 times larger simulation described in the previous section.
Notice that the mean $k_{\rm SZ}$ peak height is zero for all $\tau$, 
but the scatter around zero increases as $\tau$ increases.  
Fig.~\ref{fixedt} shows that, at fixed $\tau$, the distribution 
around the mean is well fit by a Gaussian.  To a good approximation, 
the width of the Gaussian increases linearly with $\tau$.  This is shown 
in Fig.~\ref{sigmatau}.  The distribution of the peak heights is got 
by integrating over all $\tau$:  
\begin{equation}
p(k_{\rm SZ}) = \int {\rm d}\tau\,p(\tau)\, G(k_{\rm SZ}|\tau),
\end{equation}
where $G$ denotes the Gaussian distribution of $k_{\rm SZ}$ at fixed 
$\tau$.  Because the width of the Gaussian depends strongly on $\tau$, 
the resulting $p(k_{\rm SZ})$ is a summation of Gaussians of different
dispersions.  
This is the fundamental reason why the distribution $p(k_{\rm SZ})$ is 
likely to be highly non-Gaussian.  The histogram in Fig.~\ref{kSZpdf} 
shows this distribution---note how centrally cusped it is.  

In what follows, we will describe why the distribution of $k_{\rm SZ}$
at fixed $\tau$ is Gaussian, and why $\sigma(k_{\rm SZ}|\tau)
\propto\tau$.  
We will then provide a model for $p(\tau)$.  When inserted into the 
expression above, these ingredients allow us to model the distribution 
of $k_{\rm SZ}$---this produces the solid curve shown in 
Fig.~\ref{kSZpdf}.  

\subsection{The model}
If we think of the integral along the line of sight $l$ as a sum over 
discrete pixels, then we can ask how the density and the velocity in 
pixels change as we step through $l$.  As we do this, it is likely that 
some pixels along the line of sight will contain, or lie within, 
dark matter haloes.  Because the virial motions within haloes are 
random, the component of $v(l)$ which is due to the virial motions will 
fluctuate around zero.  The result of integrating over all $l$ is that 
these internal virial motions will cancel out, so only the bulk peculiar 
velocities of the haloes contribute to the line of sight integration.  
In addition, there may be contributions to the integral from pixels
which do not contain haloes.  However, the density within a halo is on 
the order of two hundred times $\rho_{\rm crit}$.   So, provided that
the 
motions of haloes are not significantly smaller than the motions in less 
dense regions, and provided that velocities are not correlated over 
scales which are on the order of two hundred times larger than the 
size of a typical halo, pixels associated with haloes contribute much 
more to the line of sight integral than do the much less dense pixels 
which have nothing to do with haloes.  In this respect, our analysis 
is similar to that in Cole \& Kaiser (1988) and 
Peebles \& Juszkiewicz (1998).   

Haloes are rare, so, for sufficiently small pixels, it is unlikely 
that a given line of sight will contain more than one halo.  Moreover, 
the haloes have central density cusps (e.g. Navarro, Frenk \& White
1997).  Therefore, peaks in the $k_{\rm SZ}$ distribution occur in 
those pixels which contain the density cusps.  We assume that there is 
one density cusp per halo, so the number density of peaks in the 
$k_{\rm SZ}$ distribution should approximately equal the number density 
of haloes.  Moreover, we can replace the integral over the line of sight 
by the value the integrand had at the position of the peak:  
\begin{equation}
k_{\rm SZ} = \left({v\over c_{\rm light}}\right)\ 
             {\Omega_b\over\Omega_0}{\sigma_{\rm T}\over m_{\rm proton}}
             \rho_{\rm halo}\,R_{\rm halo}
           = (v/c_{\rm light})\,\tau,
\end{equation}
where $v$ is the bulk velocity of the halo, 
$c_{\rm light}$ is the speed of light, 
$\rho_{\rm halo}$ and $R_{\rm halo}$ are intended to denote the 
average density contributed by the halo over the range of 
pixels along the line of sight, $R_{\rm halo}$, it occupied.  
The final expression defines $\tau$, the optical depth of the pixel 
containing the central density cusp of the halo.   
So, to model the distribution of $k_{\rm SZ}$, we must be able to model 
how the optical depth $\tau$ and the velocity $v$ depend on halo mass 
$m$.  In addition, we must be able to estimate the number density 
$n(m)$ of haloes.  

The optical depth in a square pixel of side $L$ which contains the 
central density cusp of a halo of mass $m$ and central concentration 
parameter $c$, can be estimated by setting 
\begin{equation}
 \tau(m) = 2\,\int_0^W {{\rm d}w\over W}\,{w\over W}\,\tau(w|c,m),
 \label{taumav}
\end{equation}
where $W=[L/r_{\rm s}(m)]/\sqrt{\pi}$, and 
\begin{eqnarray}
 \tau(w|c,m) &=& {\sigma_{\rm T}\over m_{\rm proton}} 
             {\Omega_{\rm b}\over\Omega_0}\,r_{\rm s}\int {\rm d}l\
             \rho(x|c,m)
             \nonumber\\
           &=& {\sigma_{\rm T}\over m_{\rm proton}}\,
             {\Omega_{\rm b}\over \Omega_0}\, r_s(m)\rho_s(m)\,
             {2 [1-h(w)]\over w^2-1}  
\end{eqnarray}
(e.g. Cramphorn 2001).  
This expression for $\tau(w|m)$ is got by expressing all distances in 
units of $r_s(m)$ and then projecting a Navarro, Frenk \& White (1997) 
profile along the line-of-sight:  
\begin{eqnarray}
 \rho(x|c,m) &=& {\rho_{\rm s}\over x(1+x)^2},
                 \qquad {\rm where}\ x^2=l^2 + w^2, \nonumber \\
 \rho_{\rm s}(m)&=&\rho_{\rm crit}\delta_c(m)\nonumber \\
          &=&\rho_{\rm crit}{\Delta_{\rm nl}\over 3} 
              {c^3(m)\over \ln[1+c(m)]-c(m)/[1+c(m)]},\nonumber \\
 r_{\rm s}(m) &=& {r_{\rm vir}\over c(m)} = {1\over c(m)}
          \left({3m\over 4\pi\rho_{\rm crit}\Delta_{\rm
nl}}\right)^{1/3},
          \nonumber\\
 c(m) &=& 12\left({m\over m_*}\right)^{-0.13}, \nonumber \\
 h(w)  &=& {{\rm arccosh}(1/w)\over \sqrt{1-w^2}}
           \quad\mbox{if $w\le 1$} \nonumber\\
       &=& {\arccos(1/w)\over \sqrt{w^2-1}}\ \quad\mbox{if $w>1$},
\label{nfwparams}
\end{eqnarray}
and $\Delta_{\rm nl} \approx 100$ for the $\Lambda$CDM model we are 
considering here.  The quantity $c(m)$ is often called the concentration 
parameter of the halo.  Massive haloes are less centrally concentrated 
(Navarro, Frenk \& White 1997); we use the parametrization of this mass 
dependence given by Bullock et al. (2001).  
The final averaging over the circular beam of radius $W$ can also be 
done analytically:  
\begin{equation}
\tau(m) = {\sigma_{\rm T}\over m_{\rm proton}}\,
          {\Omega_{\rm b}\over \Omega_0}\, r_{\rm s}(m)\rho_{\rm s}(m)\,
           {2\over W^2}\,\Bigl[\ln\left({W^2/4}\right) + 2h(W)\Bigr].
\label{taunfw}
\end{equation}
\begin{figure}
\centering
\epsfxsize=\hsize\epsffile{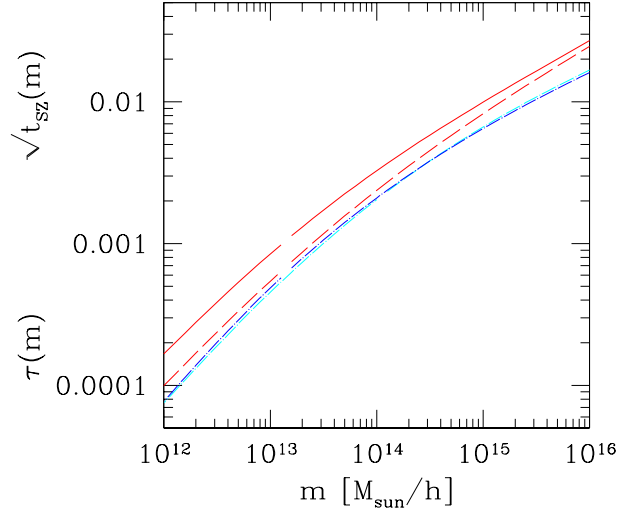}
\caption{Optical depth $\tau$ and thermal SZ effect $t_{\rm SZ}$ 
peak heights as functions of halo mass $m$.  Lower dot-dashed curves 
show $\tau(m)$ for profiles of the form given by Hernquist (1990) and 
by Navarro et al. (1997).  Upper solid and dashed curves show 
$\sqrt{|t_{\rm SZ}|}$ for Hernquist profiles when the circular 
velocity and the isotropic dispersion, respectively, are used to
estimate the temperature. } 
\label{taum}
\end{figure}

One of the dot-dashed curves in Fig.~\ref{taum} shows the optical depth 
given by equation~(\ref{taunfw}).  
Notice that, for low mass haloes, $L\gg r_{\rm s}$, and the term in
square brackets in equation (\ref{taunfw}) tends to a constant.  
In this limit $\tau\propto r_s^3\rho_s$ (recall that 
$W^2\propto r_{\rm s}^{-2}$), which means that the optical 
depth is approximately proportional to $m$.  This is sensible, since, 
in this limit, the entire halo fits in the cell.  The more massive 
haloes have larger scale radii and so they also have larger optical 
depths; at large $m$, $\tau(m)$ is approximately proportional to 
$m^{2/3}$ (see Fig.~\ref{taum}).  
For our purposes in this section, the important point is that the 
optical depth is a monotonically increasing function of halo mass.  

To show that this conclusion is not terribly sensitive to the details 
of the density run in the outer regions of haloes, the other dot-dashed 
curve shows the optical depth associated with the profile shape
presented 
in Hernquist (1990).  The Appendix shows that, for this profile shape 
also, the relevant integrals can all be computed analytically.  
(We will describe the other two curves in the next section.)  
The Hernquist profile scales as $1/x/(1+x)^3$; although it falls more 
steeply at large radii, it has the same small scale slope as the 
Navarro et al. profile.  The fact that the optical depths for these two 
parametrizations of halo profile shapes are so similar shows that most 
of the contribution to the optical depth comes from the inner parts of 
the haloes, where the profiles themselves are similar.  

What about halo speeds?  
Sheth \& Diaferio (2001) showed that, if the initial density fluctuation 
field was Gaussian, then to a good approximation, the speeds of dark 
matter haloes are drawn from a Maxwellian distribution, even at late 
times.  In addition, they showed that the rms halo speed could be
computed from linear theory, and that it is approximately independent 
of halo mass.  For example, at $z=0$ in the $\Lambda$CDM model we are 
considering, 
\begin{equation}
V_{\rm 1d}(m) = {430/\sqrt{3}\over 1 + (m/2487)^{0.284}} \ {\rm km/s},
\end{equation}
where the halo mass $m$ is expressed in units of $10^{13} M_\odot/h$.
At $z=0$ a typical halo has $m\approx 10^{13}M_\odot/h$, so that the 
results to follow are essentially unchanged if we set 
 $V_{\rm 1d} = 387/\sqrt{3}$~km/s and ignore the $m$ dependence.  
This means that, to a good approximation, the distribution of halo
speeds along the line of sight, $v$, is Gaussian, and that this Gaussian 
is independent of the optical depth along the line of sight, $\tau$.  

Since the peaks in the kinematic effect have 
 $k_{\rm SZ}\sim\tau\,(v/c_{\rm light})$, 
where $\tau$ and $v$ are the values for the halo in the 
peak pixel, the discussion above implies that the distribution of 
$k_{\rm SZ}$ peak heights is really obtained by taking the product of 
two independent distributions.  In particular, because a halo's motion 
is independent of its mass, the distribution of $v$ is independent of 
$\tau$, so the width of the distribution of $k_{\rm SZ}$ at fixed 
$\tau$ should be proportional to $\tau$.  This is consistent with our 
Fig.~\ref{sigmatau}.  Secondly, at fixed $\tau$, the distribution of 
$k_{\rm SZ}$ is really just the distribution of a constant times the 
distribution of $v$.  Because halo motions are Gaussian, the
distribution 
of $k_{\rm SZ}$ at fixed $\tau$ should be Gaussian; this is consistent 
with our Fig.~\ref{fixedt}.  (It is worth adding that 
Sheth \& Diaferio 2001 showed that the rms motions of haloes are higher 
in the dense regions.  This  explains why, in the lower panels in 
Figure~3 of Diaferio et al. 2000, the scatter in $k_{\rm SZ}$ at fixed 
$\tau$ is larger in the densest regions of their simulations.)

Finally, because peaks correspond to haloes, the comoving number density 
of peaks equals the comoving number density of haloes.  
Let $p(\tau|m)\,{\rm d}\tau$ denote the probability that a halo of mass 
$m$ has optical depth $\tau$.  Then 
\begin{eqnarray}
 p(k_{\rm SZ})\,{\rm d}k_{\rm SZ} 
           &=& \int {{\rm d}m\, n(m)\over n_{\rm haloes}}\,
               \int {\rm d}\tau\,p(\tau|m)\ \times \nonumber \\
     &&\qquad\qquad
           {p(v/c_{\rm l}=k_{\rm SZ}/\tau|\tau,m)\,{\rm d}k_{\rm
SZ}\over \tau}
 \nonumber \\
 &=& \int {\rm d}m\, p(m)\,\int {\rm d}\tau\,p(\tau|m)\ \times
\nonumber\\
 && \qquad\qquad 
    {p(v/c_{\rm l}=k_{\rm SZ}/\tau)\,{\rm d}k_{\rm SZ}\over \tau}
\nonumber \\
  &=& \int {\rm d}\tau\, 
      {p(v/c_{\rm l}=k_{\rm SZ}/\tau)\,{\rm d}k_{\rm SZ}\over \tau}\,
      p(\tau) \ \times \nonumber \\
  && \qquad \qquad \int {\rm d}m\,p(m|\tau),
 \label{pkSZfull}
\end{eqnarray}
where we have written $c_{\rm l}$ instead of $c_{\rm light}$, 
and $n_{\rm haloes} \equiv \int {\rm d}\tau\, n(\tau)$.  
The second equality follows from writing 
 $n(m)/n_{\rm haloes}\equiv p(m)$, 
and using our assumption that the distribution of $v$ is approximately
independent of halo mass.  The final expression follows from using the 
fact that $p(m)\,p(\tau|m) = p(m|\tau)\,p(\tau)$ and rearranging the 
order of the integrals.  
Because the integral over $m$ leaves a function of $\tau$ only, 
the final expression shows that, at fixed $\tau$, the distribution of 
$k_{\rm SZ}$ is given by the distribution of $v$.  Since this is 
Gaussian, our model produces a distribution of $k_{\rm SZ}$ which, 
at fixed $\tau$, is Gaussian, in agreement with the simulations 
(Fig.~\ref{fixedt}).  

To compute our model for the full distribution of $k_{\rm SZ}$, we 
need a model for the distribution of $\tau$ at fixed $m$.  We will 
do this in the next section.  For now, note that if this distribution 
is sharply peaked about a mean value, say $\tau(m)$ given by 
equation~(\ref{taumav}), then we can replace $p(\tau|m)$ with a 
delta function.  In this case, 
\begin{equation}
p(k_{\rm SZ}) = \int {{\rm d}m\, n(m)\over n_{\rm haloes}}\,
 {{\rm e}^{-[k_{\rm SZ}/\tau(m)/(V_{\rm 1d}/c_{\rm l})]^2/2}
  \over\sqrt{2\pi}(V_{\rm 1d}/c_{\rm l})\tau(m)},
\label{pkSZ}
\end{equation}
where we have explicitly shown what happens when we substitute the 
Gaussian with width $V_{\rm 1d}$ for the distribution of halo
velocities.  

The final integrand above is the product of the halo mass function 
and a Gaussian whose width increases as $m$ increases.  This sort of 
integral is just like that studied by Sheth \& Diaferio (2001) in their 
model of why the peculiar velocity distribution of dark matter particles 
is not Maxwellian.  In that case, the final distribution was got by 
summing Gaussians of different widths because the virial velocities did 
contribute to the statistic, and the rms virial velocities within haloes 
scale as $m^{1/3}$.  Here, the virial motions do not enter, but the 
Gaussians have different widths because the optical depth scales 
approximately as $\tau(m) \propto m^{2/3}$.  

In the next section, we will show that, in fact, $p(\tau|m)$ is not a 
delta function.  Nevertheless, it is reasonably sharply peaked, so that 
the delta function is not a bad approximation.  Of course, there is a 
small price to pay for the gain in simplicity.  Our final expression 
above suggests that the distribution of $k_{\rm SZ}$ at fixed $m$ 
should be Gaussian.  This is a direct consequence of our neglect of 
the scatter in $\tau$ at fixed $m$.  If we include the effects of the
scatter (along the lines we describe in the next section) then the first 
line of equation~(\ref{pkSZfull}) above shows clearly that, even at
fixed 
$m$, the distribution of $k_{\rm SZ}$ is got by summing up Gaussians of 
different dispersions, so it will be quite non-Gaussian.  

Finally, we need to specify how the number density of haloes depends 
on halo mass.  To get the solid curve in Fig.~\ref{kSZpdf}, we used 
the halo mass function $n(m)$ given by Sheth \& Tormen (1999).  The 
model provides a good description of the simulations at $z=0$.  The 
dashed curves show the predicted contribution to the distribution from
haloes 
in different mass bins:  the central spike of the distribution is 
from small mass haloes ($10^{12}-10^{13}M_\odot/h$), whereas the 
broad wings are entirely due to the more massive haloes 
($\ge 10^{14}M_\odot/h$).  

\section{The thermal effect}\label{tszeffect}
The CMB temperature fluctuation due to the thermal effect is 
\begin{eqnarray}
t_{\rm SZ} &=& {\Delta T\over T} = 
\left(x\frac{e^{x}+1}{e^{x}-1}-4\right) \int n_{\rm e}\,\sigma_{\rm T}\,
{k_{\rm B}T_{\rm e}\over m_{\rm e}\,c_{\rm light}^2}\ {\rm d}l \nonumber
\\
    &=&-2 \int n_{\rm e}\,\sigma_{\rm T}\,
          {k_{\rm B}T_{\rm e}\over m_{\rm e}\,c_{\rm light}^2}\ {\rm d}l 
           \qquad\qquad ({\rm as}\ \ x\to 0)\nonumber \\
    &\approx& -2\,{\sigma_{\rm T}\over m_{\rm e}} 
       {\Omega_{\rm b}\over\Omega_0}\int {\rm d}l\ 
       {\sigma^2_{\rm dm}(l)\over 2c_{\rm light}^2}\,\rho_{\rm dm}(l),
\label{tSZl}
\end{eqnarray}
where $x\equiv h\nu/kT$, the second line shows what happens in the 
Rayleigh-Jeans limit ($x\to 0$, the only limit we will consider in 
this paper), and the third line shows the 
approximation Diaferio et al. (2000) used to relate the density 
$\rho_{\rm dm}$ and line-of-sight velocity dispersion 
$\sigma^2_{\rm dm}$ of dark matter particles in their simulations to 
the density and temperature of electrons.  Namely, we are assuming that 
the gas and the dark matter have similar spherically symmetric density 
profiles with isotropic velocity dispersions, so the additional
assumption 
of hydrostatic equilibrium determines the velocity dispersion profile of 
the gas uniquely.

Before moving on, a word on the relative amplitudes of the thermal and 
kinematic effects is in order.  
The first line of equation~(\ref{tSZl}) shows that the amplitude of 
the thermal effect depends on frequency.  The kinematic effect, on 
the other hand, is independent of frequency.  
In the Rayleigh-Jeans limit, the amplitude of the thermal effect is 
much larger than that of the kinematic effect.  However, the thermal 
effect effectively vanishes at a frequency near 217GHz.  Because 
the kinematic effect does not depend on frequency, the non-Gaussian 
signature due to the kinematic effect is important near 217GHz.

\begin{figure}
\centering
\epsfxsize=\hsize\epsffile{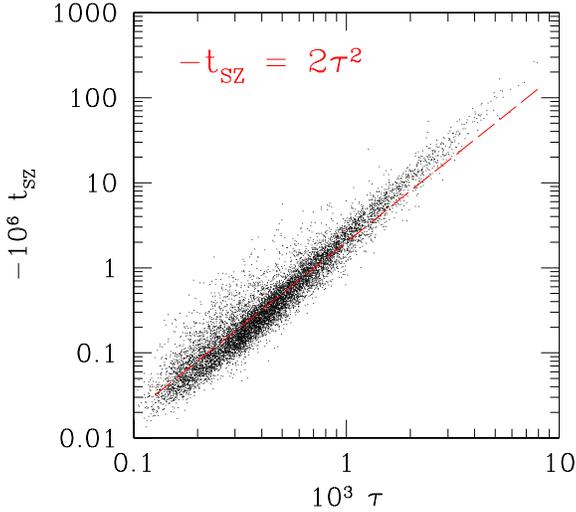}
\caption{Scatter plot showing the height of a $t_{\rm SZ}$ peak 
as a function of optical depth $\tau$.}
\label{tscat}
\end{figure}

Diaferio et al. showed that, to a good approximation, 
$t_{\rm SZ}\approx -2\tau^2$, where the optical depth $\tau$ was 
defined in the previous section.  Fig.~\ref{tscat} shows that this 
relation also describes the thermal effect peaks in our 
considerably larger simulation box.  
This section describes a model for why this happens, which allows 
us to compute the distribution of $t_{\rm SZ}$ peak heights.  

\subsection{The mean relation between $\tau$ and $t_{\rm SZ}$}

Our model for the distribution of peaks in the thermal effect is similar 
in spirit to that for the kinematic effect:  each peak in $t_{\rm SZ}$ 
is associated with the centre of a dark matter halo, we ignore the
effect of haloes overlapping along the line of sight.
This means that we can replace the 
integral over the line of sight in equation~(\ref{tSZl}) by an integral 
over the density-weighted temperature profile of the single halo of mass 
$m$ and concentration $c$ which happened to be in the line of sight:  
\begin{eqnarray}
t_{\rm SZ}(c,m) &=& -2r_{\rm s}\,
   {\sigma_{\rm T}\over m_{\rm e}} {\Omega_{\rm b}\over\Omega_0}
   \int_0^W {2w\,{\rm d}w\over W^2} \times \nonumber\\
&& \qquad\qquad 
   \int {\rm d}l\ {\sigma^2(x|c,m)\over 2c_{\rm light}^2}\,\rho(x|c,m),
\end{eqnarray}
where $x^2 = l^2 + w^2$, and all distances are in units of the scale 
radius $r_{\rm s}=r_{\rm vir}/c(m)$, with $r_{\rm vir}$ the virial 
radius and $c$ the concentration parameter (equation~\ref{nfwparams}).  
As before, the integral over $w$ represents the beam smoothing.  
This shows that the thermal effect is proportional to the integral of 
the density times the internal velocity dispersion, projected along 
the line of sight (recall that the kinematic effect is proportional 
to the density times the line-of-sight bulk motion of the halo).  

In principle, there is some freedom in deciding what to use for the 
velocity dispersion $\sigma^2$ within a halo.  Because we are assuming, 
as Diaferio et al. (2000) did, that the gas and the dark matter have 
similar spherically symmetric density profiles with isotropic orbits, 
the additional assumption of hydrostatic equilibrium determines the gas 
velocity dispersion profile, and hence the gas temperature profile 
uniquely.  For the Navarro et al. (1997) profile, as well as for the 
Hernquist (1990) profile which we discussed earlier, the velocity 
dispersion profile $\sigma^2(r)$ has a fairly complicated functional 
form (Hernquist 1990; Cole \& Lacey 1996).  Although the shape of 
$\sigma^2(r)$ is different from the circular velocity profile
$Gm(<r)/2r$, 
the two profiles are not very different.  The circular velocity profile 
has a much simpler functional form, and Cramphorn (2001) shows the 
result of setting $\sigma^2(r) = Gm(<r)/2r$, and then evaluating the 
required integral above for the Navarro et al. (1997) profile
numerically.  
He states that the result is well approximated by  
$\tau^2\approx -t_{\rm SZ}/2$, in agreement with the simulation 
result in Diaferio et al. 

For the Hernquist (1990) profile we discussed earlier, the integral 
above can be done analytically, both for the correct one-dimensional 
dispersion, and when the circular velocity is used to approximate the 
dispersion (see Appendix A).  
Recall that, on small scales, Hernquist's profile has the same slope as 
that of Navarro et al. (1997), so we expect that using it provides a 
reasonable analytic approximation.  The upper solid curve in 
Fig.~\ref{taum} shows our analytic expression for the beam averaged 
$\sqrt{-t_{\rm SZ}}$ associated with Hernquist profiles when the squared 
circular velocity is used for $\sigma^2$, and the dashed curve shows the 
result of using the actual dispersion.  Recall that the other curves in 
the Figure show the optical depth.  Thus, Fig.~\ref{taum} shows that 
both $|t_{\rm SZ}|$ and $\tau$ increase monotonically with halo mass, 
and that, to a good approximation, $-t_{\rm SZ}\propto \tau^2$.  
Moreover, the dashed line shows a slope slightly steeper than
two at high masses, as it is indeed seen in Fig 6.
The constant of proportionality is of order two, consistent with the 
simulations.  The difference between the solid and the dashed 
$t_{\rm SZ}$ curves can be thought of approximately illustrating what 
might happen if our assumption that the gas traces the spherically 
symmetric dark matter distribution is relaxed.  

\subsection{The distribution of $\tau$ and $t_{\rm SZ}$}
The previous subsection showed how to estimate the beam averaged 
optical depth and thermal SZ effect due to single haloes of mass $m$.  
If all haloes of a given mass had exactly the same density and velocity 
dispersion profiles, then we would be able to translate the distribution 
of halo masses into distributions of $\tau$ and $t_{\rm SZ}$.  In fact, 
haloes of fixed $m$ do not all have the same profile shape:  the 
distribution of shapes is well parametrized by assuming that the 
profile is always of the form given by Navarro et al. (1997), but 
letting the concentration parameter $c$ defined earlier follow 
a lognormal distribution (Jing 2000; Bullock et al. 2001).  
Therefore, we will assume that 
\begin{equation}
  p(t_{\rm SZ}|m)\,{\rm d}t_{\rm SZ} = 
 {{\rm d}c/c\over \sqrt{2\pi\sigma^2_{\rm c}}}
\exp\left(-{\ln^2[c(t_{\rm SZ},m)/\bar c(m)]\over 2\sigma^2_{\rm
c}}\right) ,
\end{equation}
where the mean concentration at fixed mass, $\bar c(m)$, is
given by equation~(\ref{nfwparams}), and the rms scatter, 
$\sigma_{\rm c}\approx 0.2$, is approximately independent of halo 
mass.  The number density of the thermal effect peaks of height 
$t_{\rm SZ}$ is just this times the number of haloes of mass $m$, 
integrated over all $m$:
\begin{equation}
  n(t_{\rm SZ})\,{\rm d}t_{\rm SZ} = 
  \int {\rm d}m\ n(m)\,p(t_{\rm SZ}|m)\,{\rm d}t_{\rm SZ}.
\end{equation}
To proceed, we need $c(t_{\rm SZ},m)$.  This can be got by inverting 
the $t_{\rm SZ}(c,m)$ relation given above.  
Recall that $t_{\rm SZ}(c,m)$ can be evaluated analytically 
if the profile has the form given by Hernquist (1990), so that 
the Jacobian ${\rm d}c/{\rm d}t_{\rm SZ}$ can also be given
analytically.  

The number density of peaks of height $\tau$ in the optical depth 
distribution can be got analogously:  
\begin{equation}
  n(\tau)\,{\rm d}\tau = \int {\rm d}m\ n(m)\,p\Bigl(c(\tau,m)|m\Bigr)
                   \,{{\rm d}c(\tau,m)\over {\rm d}\tau}\,{\rm d}\tau.
\end{equation}

\begin{figure}
\centering
\epsfxsize=\hsize\epsffile{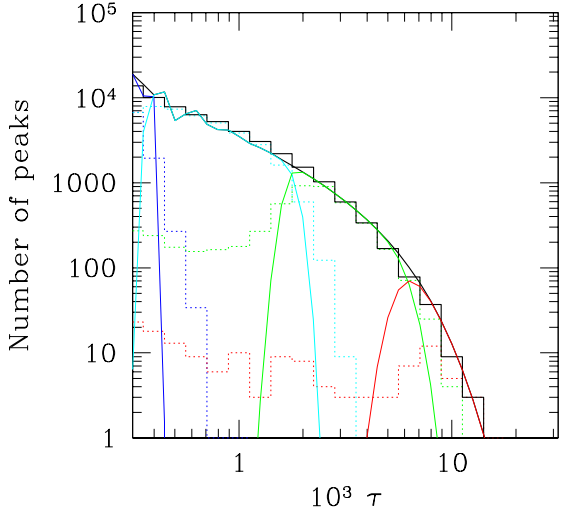}
\caption{Distribution of optical depth peaks $\tau$ in the 
simulations (solid histograms) and in our model (smooth solid curves).  
Dashed lines show the contributions to the total from 
haloes with mass in the range $10^{13}$--$10^{14}$, 
$10^{14}$--$10^{15}$, and $\ge 10^{15}M_\odot/h$.} 
\label{taupdf}
\end{figure}

\begin{figure}
\centering
\epsfxsize=\hsize\epsffile{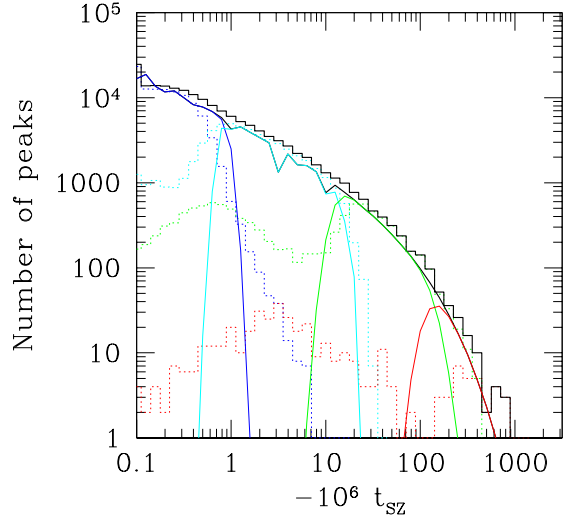}
\caption{Distribution of peak heights $t_{\rm SZ}$ in the thermal effect 
in the simulations (solid histograms) and in our model (smooth solid 
curves).  Both simulations and model assume that the gas traces the 
dark matter.  Dashed lines show the contributions to 
the total from haloes with mass in the range $10^{13}$--$10^{14}$ 
(dominate at small $\tau$), $10^{14}$--$10^{15}$, and 
$\ge 10^{15}M_\odot/h$ (dominate at large $\tau$).} 
\label{tSZpdf}
\end{figure}

Figs.~\ref{taupdf} and~\ref{tSZpdf} show the distributions of 
optical depth and thermal effect peak heights in the simulations.  
Solid lines show the total number density of peaks, and dashed lines show the contribution to the total from haloes in 
small mass ranges. In the model, for the sake of simplicity, 
we compute $t_{\rm SZ}$ (see equation (13)) using the
Hernquist profile and its corresponding circular velocity.
The dashed lines were computed by locating dark 
matter haloes using the spherical overdensity algorithm with density 
threshold 200 (details are in Tormen 2001, in preparation).  Simulation 
particles which reside in the dark matter haloes in a given mass range 
were marked, and then the SZ effect and Thomson optical depth were 
computed by using only the marked particles.

These distributions are bimodal; our model fits the large $\tau$ 
peak for each mass range reasonably well.  This suggests that the large 
$\tau$ peak is due to the central cusp of a halo, whereas the increase 
at small $\tau$ is primarly due to the substructure of haloes which 
our model does not account for.  Note, however, that by the time 
we are seeing the effect of halo-substructure, the dominant 
contribution to $t_{\rm SZ}$ is from the central cusps of less 
massive haloes.  

The careful reader will have noticed that a scatter in $c$ at fixed $m$ 
leads to a scatter in optical depth at fixed mass $\tau(m)$.  
This, in turn, leads to scatter in the kinematic effect $k_{\rm SZ}$, 
which is in addition to the scatter induced by the fact that haloes 
are moving with different speeds $v$.  In ignoring the scatter which 
is due to the distribution of halo concentrations at fixed $m$,  
we, in effect, assumed that the scatter in bulk velocities $v$ was 
the dominant cause of the scatter in $k_{\rm SZ}$ at fixed $m$,
as it is indeed.

\section{Discussion}
We presented a simple model for the distribution of peak heights 
in maps of the kinematic and thermal Sunyaev-Zeldovich effects.  
In our model, which is similar in spirit to that proposed by 
Cole \& Kaiser (1988), there is a one-to-one correspondence between 
a peak in the kinematic or the thermal effect and the presence of a 
massive cluster.  So the shape of the distribution of peaks is
determined 
by the mass function of clusters, and by how clusters move.  
If the distribution of initial density fluctuations was Gaussian, 
then the motions of clusters along the line of sight should be
reasonably 
well fit by a Gaussian, so one might have expected the distribution of 
$k_{\rm SZ}$ peak heights to also be Gaussian.  
Our simulations showed that, in fact, the distribution of $k_{\rm SZ}$ 
peak heights is highly non-Gaussian.  We argued that this happens
because 
the kinematic effect is proportional to the product of the cluster 
velocity and its optical depth, the optical depths of clusters depend 
strongly on cluster mass, and the mass range of clusters which cause 
peaks in the kinematic effect is quite large.  In our model, the 
cluster mass function and the rms motions of clusters are quantities 
which can be computed if the initial spectrum of density perturbations 
is specified.  Therefore, the kinematic effect can be used to constrain 
the shape of this spectrum.  

\begin{figure}
\centering
\epsfxsize=\hsize\epsffile{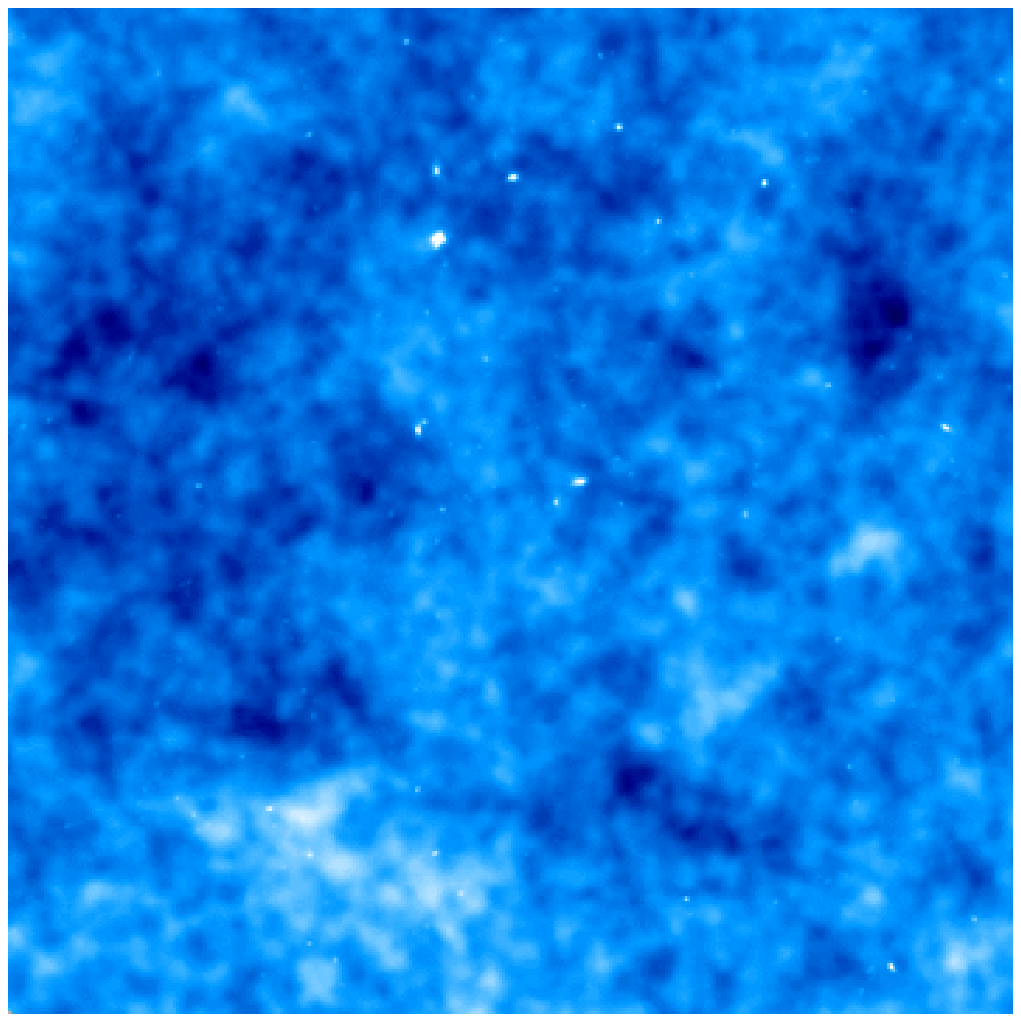}
\vspace{2mm}
\epsfxsize=\hsize\epsffile{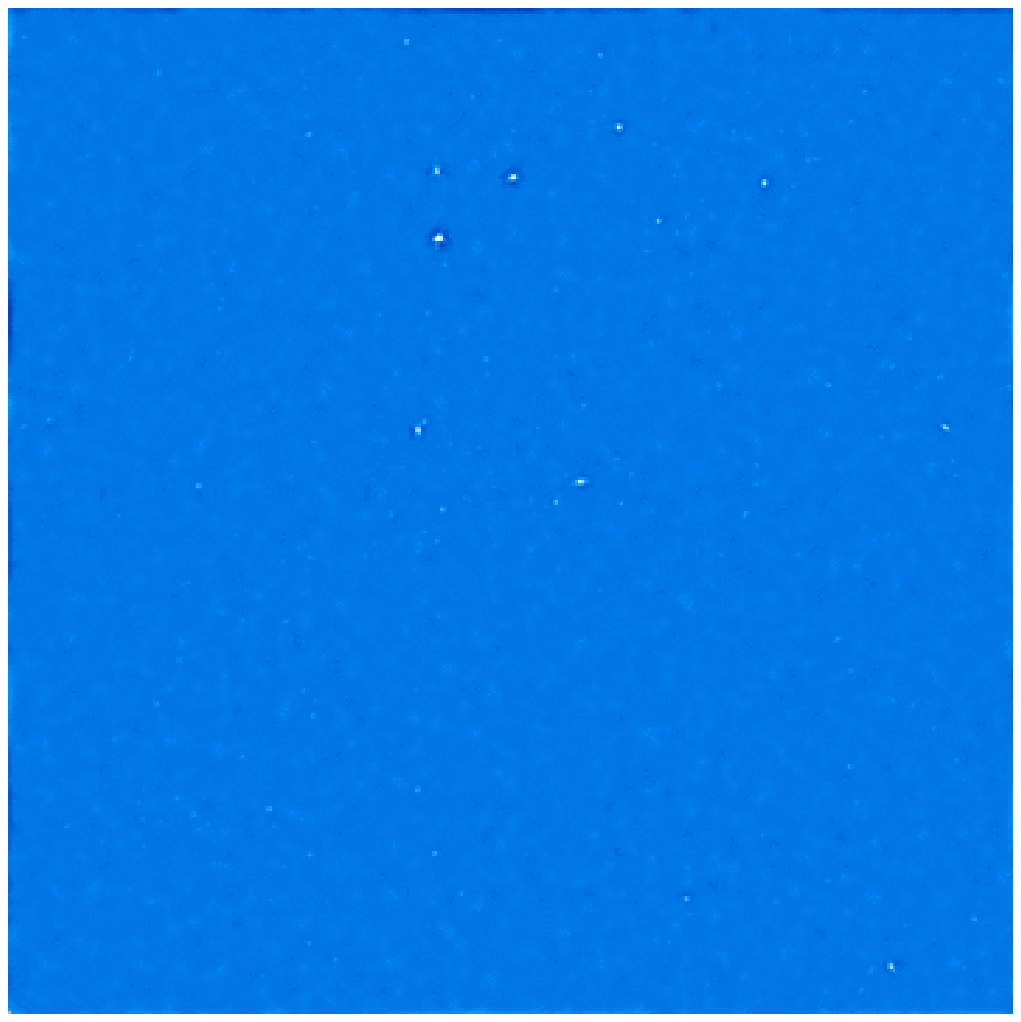}
\caption{Distribution of temperatures in simulated 
 8$^\circ$ $\times$ 8$^\circ$ maps of the intrinsic CMB fluctuations 
with the thermal SZ effect fluctuations superimposed.  
Top panel shows the full map, and bottom panel shows what remains 
after applying a high-pass filter of scale one arcminute.
The maps shown here are part of the full map computed from the 
entire 480Mpc$/h$ box.} 
\label{maps}
\end{figure}

Our model also allowed us to estimate the distribution of peak heights 
in the optical depth (which is not observable) and in the thermal effect
(which is).  To illustrate the logic, we showed what one predicts for 
the shapes of these non-Gaussian distributions if gas traces the dark 
matter.  Our model was able to describe the simulations quite well.  
For the optical depth and for the thermal effect, the non-Gaussianity 
was a consequence of the non-Gaussian shape of the halo mass function.  
The main aim of this paper was to show how our knowledge of halo speeds 
and profiles can be used to model the thermal and kinematic SZ effects.  

Real CMB maps will have the peaks of the SZ effects on top of the 
primary fluctuations. The primary fluctuations are Gaussian distributed 
with an rms, on arcminute scales, of about $4\times 10^{-5}$.  
This rms is a few times larger than the typical thermal SZ effect, 
and an order of magnitude larger than the typical kinematic SZ effect. 
However, in our $\Lambda$CDM model, the SZ effect fluctuations and 
the primary fluctuations dominate the angular power spectrum of the CMB 
on substantially different scales, which separate at $\ell\sim 3000$ 
(e.g. Springel, White \& Hernquist 2001).  
Thus, in principle, an optimal high-pass filter for the arcminute scale 
should be able to isolate the peaks due to the SZ effects from the
primary fluctuations. 

\begin{figure}
\centering
\epsfxsize=\hsize\epsffile{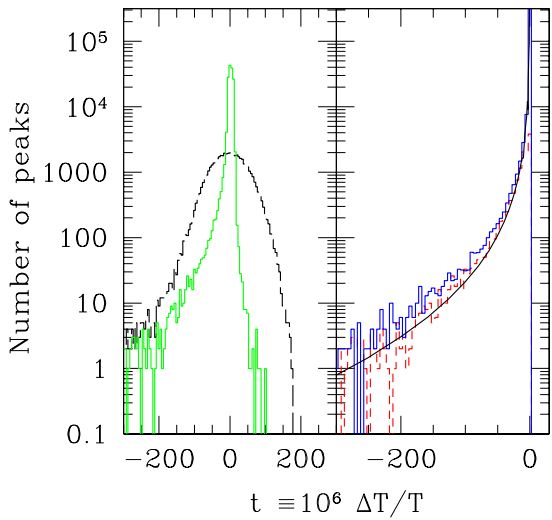}
\caption{Distribution of peak heights in simulated temperature maps.  
Dashed histogram in panel on left shows the peaks in the unfiltered 
map shown (Fig.~\ref{maps} top).  Solid, slightly asymmetric histogram 
shows the peak heights in the high-pass filtered map (Fig.~\ref{maps} 
bottom).  Right panel compares the actual $t_{\rm SZ}$ peak decrements
(solid histogram) and the peak distribution after subtracting the 
positive peak height distribution from the negative side in the filtered 
map (dashed histogram). Solid curve shows what our model predicts.  
The final distribution in the filtered map is similar to the actual 
$t_{\rm SZ}$ distribution, and is similar to what our model predicts.} 
\label{filtpdf}
\end{figure}

If we put our simulation box at z=0.218, then the 
angular size of one pixel is 1 arcminute, and the 
whole box occupies 40$^\circ$ $\times$ 40$^\circ$ 
square-degrees.  In Fig.~\ref{maps}, we show a small 
8$^\circ$ $\times$ 8$^\circ$ piece of the whole map, 
so as to make smaller angular scale structures visible. 
For the analysis in Fig.~\ref{filtpdf}, however, we use the entire 
map. 
The top panel in Figure~\ref{maps} shows a simulated temperature map in 
which the $t_{\rm SZ}$ effect fluctuations are superimposed on the
primary fluctuations generated at the last scattering epoch. We made a two
dimensional random Gaussian field for the primary temperature
fluctuations. We computed the angular power spectrum by CMBFAST for the
$\Lambda$CDM model we consider here.
The bottom panel shows the effect of applying a high-pass filter 
of scale one arcminute to this map; the large scale modulations are
gone, and only small scale fluctuations remain. 

Fig.~\ref{filtpdf} shows the distribution of temperature peak heights 
in the simulated map which has both primary and $t_{\rm SZ}$ fluctuations 
(dashed histogram) and in the high-pass filtered map (solid histogram); 
i.e., the peak height distributions in the maps shown in Fig.~\ref{maps}.  
The Gaussian-like shape of the dashed curve for the unfiltered map is 
primarily due to the peak height distribution of the intrinsic peaks 
(we have verified that it is similar to the analytic formula for peaks in 
two-dimensional Gaussian random fields given in Bond \& Efstathiou 1987); 
the $t_{\rm SZ}$ peaks account for most of the large decrement tail.  
The solid histogram in the left panel, from the filtered map, is slightly 
asymmetric.  If we assume that the positive peaks are due to the intrinsic 
CMB fluctuations, whereas the decrements are either intrinsic or 
$t_{\rm SZ}$ peaks, then subtracting the positive side from the negative 
one should leave the $t_{\rm SZ}$ peak distribution.  The dashed histogram 
in the right panel shows what remains after doing this subtraction.  
It should be compared with the true distribution of $t_{\rm SZ}$ peaks 
which we presented earlier in the paper, and is now shown as the solid 
histogram.  The smooth solid curve shows what our model predicts.
Notice how similar the dashed and solid histograms are.  
This demonstrates that by suitably filtering the map, it should be 
possible to recover the true distribution of $t_{\rm SZ}$ peaks, and 
that the our model does a good job of describing this distribution. 

Unfortunately, this method will not work for recovering the distribution 
of the $k_{\rm SZ}$ peaks.  To distinguish the kinematic from the
thermal effect, we must use their different spectral properties: 
multi-band follow-up observations of regions with deep and highly 
clustered $t_{\rm SZ}$ decrements will assure the presence of a 
supercluster where the $k_{\rm SZ}$ effect is largest 
(e.g. Diaferio et al. 2000).      

In fact, the $t_{\rm SZ}$ peaks will also show up in clustering 
statistics.  The shape of the correlation function of the peaks which 
were imprinted on the background radiation at the last scattering
surface depends on peak height (Bond \& Efstathiou 1987; 
Heavens \& Sheth 1999; Heavens \& Gupta 2001).  The presence of 
$t_{\rm SZ}$ peaks will change this dependence because, as our model 
shows, the two-point correlation function of the SZ peaks is related 
to the two-point correlation function of massive clusters.  
Since accurate analytical models for the clustering of clusters exist 
(e.g. Mo \& White 1996; Sheth \& Tormen 1999), the clustering of the 
peaks in the SZ effect can be estimated rather easily, although we have 
not done so here.  The clustering of clusters depends differently on 
the initial spectrum of fluctuations than the clustering of intrinsic 
peaks does; therefore, our model for the $t_{\rm SZ}$ peak distribution 
will be important if one wishes to obtain constraints on the 
initial fluctuation spectrum by studying the two-point and higher order 
moments of peaks in the microwave background on arcminute scales.  

Finally, note that our model assumes that there is a one-to-one 
correspondence between a peak in the kinematic or thermal effects and 
the presence of a massive cluster.  We showed that, at low redshift, 
this is a good approximation--our model describes the simulations quite 
well.  At higher redshifts this assumption is likely to break down
for the kinematic SZ effect.  
This is because, at higher redshifts, there are fewer massive haloes 
with sufficiently high optical depths to produce obvious peaks.  In 
particular, the number density of massive clusters drops faster than 
does the typical coherent--flow speed, so that, at higher redshifts, an 
increasing fraction of peaks will be caused by the velocity field,
rather than by the density field.  

Since the kinematic peaks due to coherent flows at high redshift have 
a different angular structure from those due to nearby clusters, an 
optimal filter will be able to distinguish the two types of peaks. 
Developing a model for this additional effect is the subject of work 
in progress.

\bigskip

We thank Simon White, Rashid Sunyaev, Naoshi Sugiyama, Atsushi Taruya
and Adi Nusser for helpful comments.  This research was partly supported 
by the NATO Collaborative Linkage Grant PST.CLG.976902.  
RKS is supported by the DOE and NASA grant NAG 5-7092 at Fermilab.

\appendix 

\section{Projections of Hernquist profiles}
The main text uses expressions for the density and the density-weighted 
temperature profiles of dark matter haloes integrated along the line of 
sight.  For the Hernquist profile, both these integrals can be done 
analytically (Hernquist 1990).  (Analytic results can also be obtained 
for many of the more general profiles in Zhao 1996; in the interests 
of brevity, we have not provided explicit expressions here.)  

The density at $x = r/r_{\rm s}$, where $r_{\rm s}$ is a scale radius, 
from the centre of a Hernquist profile is 
\begin{equation}
\rho(x) = {m\,(1+b)^2\over 2\pi\, b^3r_{\rm vir}^3\, x(1+x)^3},
\end{equation}
where $b=r_{\rm s}/r_{\rm vir}$ is the scale radius in units of the
virial 
radius.  Reasonable agreement with Navarro et al. (1997) profiles 
of the same $m$ and concentration $c$ can be got by setting 
$b=\sqrt{2}/c^{0.75}$ (Sheth et al. 2001).

The integral over the density, which is related to the optical 
depth, is 
\begin{equation}
 I(z) = \int \rho(x)\,{\rm d}l = {m\,(1+b)^2\over 2\pi\, b^2r^2_{\rm
vir}}
         {(2 + z^2)\,h(z) - 3\over (z^2-1)^2} ,
\end{equation}
where, in the integrand, $x^2 = l^2 + z^2$, with $x$, $l$, and $z$ all 
in units of the scale radius $br_{\rm vir}$, and $h(z)$ is given in the 
main text.  This quantity, averaged over a circular window of radius
$R$, 
is 
\begin{equation}
 2\int_0^Z {{\rm d}z\over Z}{z\over Z}\ I(z) = 2\,
 {m\,(1+b)^2\over 2\pi\,b^2r^2_{\rm vir}}\, {1 - h(Z)\over Z^2 - 1},
\end{equation}
where $Z \equiv R/(br_{\rm vir})$.

There is some freedom as to how we should estimate the temperature.  
If we assume that the halo is not rotating, then the quantity of 
interest is the density-weighted line-of-sight velocity dispersion.  
If the distribution of orbits of dark matter is isotropic, and the gas moves 
like the dark matter, then this is 
\begin{eqnarray}
S(z) &=& {Gm\over br_{\rm vir}}\, 
{m\,(1+b)^4\over 2\pi\,b^2r_{\rm vir}^2}\,
\Biggl[{6 - 65z^2 + 68z^4 - 24z^6\over 12(1-z^2)^3} - \pi z \nonumber\\
&&\qquad
-{z^2 h(z)(8z^6 - 28 z^4 + 35 z^2 - 20)\over 4(1-z^2)^3}\Biggr],
\end{eqnarray}
and the average over a circle is 
\begin{eqnarray}
2\int_0^Z {{\rm d}z\over Z}{z\over Z}\ S(z) 
&=& 2\,{Gm\over br_{\rm vir}}\,{m\,(1+b)^4\over 2\pi\, b^2r_{\rm
vir}^2}\nonumber \\
&&\ \times\ 
  \Biggl[{3 - 17 Z^2 + 22 Z^4 - 8 Z^6\over 12(1 - Z^2)^3}- {\pi Z\over
3}\nonumber \\
&& \ \ + \ \ \ h(Z)\,{15 Z^2 - 20 Z^4 + 8 Z^6\over 12(1 -
Z^2)^2}\Biggr].
\end{eqnarray}
If, instead, we use one half of the circular velocity squared, 
\begin{equation}
{v_{\rm c}^2\over 2} \equiv {Gm(<r)\over 2r} 
      = {Gm\,(1+b)^2\over 2\,br_{\rm vir}} {x \over (1+x)^2},
\end{equation}
then we need 
\begin{equation}
V(z) = \int {v_{\rm c}^2(x)\over 2}\,\rho(x)\,{\rm d}l = 
 {Gm\over br_{\rm vir}}\,{m\,(1+b)^4\over 2\pi\, b^2r_{\rm vir}^2}\, 
 {h_2(z)/2\over (z^2 - 1)^4},
\end{equation}
where 
\begin{displaymath}
 h_2(z) = {6 + 83 z^2 + 16 z^4\over 12} - {5 z^2\,(4 + 3 z^2)\over
4}\,h(z).
\end{displaymath}
This, averaged over a circle of radius $Z$ is 
\begin{eqnarray}
2\int_0^Z\!{{\rm d}z\over Z}{z\over Z}\ \!V(z)\!%
&=&\!{Gm\over br_{\rm vir}}\,
       {m\,(1+b)^4\over 2\pi\, b^2r_{\rm vir}^2}
\times \nonumber\\
&\Biggl[&\!\!\!\!\!\!{-3 - 14 Z^2 + 2 Z^4\over 12(Z^2-1)^3} 
                         + {15 Z^2\,h(Z)\over 12(Z^2-1)^3}\Biggr].
\end{eqnarray}

\end{document}